\documentclass[runningheads]{llncs}
\usepackage{graphicx}
\usepackage{color}
\usepackage{tcolorbox} 
\usepackage{mathtools}
\usepackage{amsmath}
\usepackage{centernot}
\usepackage{booktabs}
\usepackage{tabularx}
\tcbuselibrary{theorems}
\usepackage{fontawesome}

\newtcbtheorem{takeaway}{\faLightbulbO{} Takeaway}
{colback=green!5,colframe=green!35!black,fonttitle=\bfseries}{th}
\newtcbtheorem{defs}{\faInfoCircle{} Definition}
{colback=red!5,colframe=red!35!black,fonttitle=\bfseries}{th}

\begin{document}
\title{7. Automated Requirements Relations Extraction}
%
%
\author{Quim Motger \and
Xavier Franch}
%
%

\institute{Department of Service and Information System Engineering, \\Universitat Politècnica de Catalunya\\ \email{\{joaquim.motger,xavier.franch\}@upc.edu}}
\maketitle              
%

%
%
%

\section*{Abstract}
In the context of requirements engineering, relation extraction involves identifying and documenting the associations between different requirements artefacts. 
When dealing with textual requirements (i.e., requirements expressed using natural language), relation extraction becomes a cognitively challenging task, especially in terms of ambiguity 
and required effort from domain-experts. 
Hence, in highly-adaptive, large-scale environments, effective and efficient automated relation extraction using natural language processing techniques becomes essential. 
In this chapter, we present a comprehensive overview of natural language-based relation extraction from text-based requirements. We initially describe the fundamentals of requirements relations based on the most relevant literature in the field, including the most common requirements relations types. The core of the chapter is composed by two main sections: (i) natural language techniques for the identification and categorization of requirements relations (i.e., syntactic vs. semantic techniques), and (ii) information extraction methods for the task of relation extraction (i.e., retrieval-based vs. machine learning-based methods). We complement this analysis with the state-of-the-art challenges and the envisioned future research directions. Overall, this chapter aims at providing a clear perspective on the theoretical and practical fundamentals in the field of natural language-based relation extraction.


\section{Introduction}
\label{sec:introduction}

In the context of software development, requirements do not exist in isolation, but often affect or are affected by the lifecycle of other requirements from the same component, system, product or project~\cite{Carlshamre2001}. 
This relational environment 
underscores the need for effective extraction – crucial for informed decisions, prioritization, and strategic planning \cite{Achimugu2014,Svahnberg2010}.
For instance, a requirement might require the completion of another requirement before its fulfilment, which makes it a key factor in determining project scope and scheduling. In contrast, a requirement might refine or elaborate further details on another, enabling a deeper understanding of its development. Moreover, multiple requirements can contain similar or even duplicate content, highlighting redundant specifications which can lead to ambiguity, inefficiency and potential conflicts in development and testing phases.
Ignoring these connections can harm project development trajectory, affecting design decisions and project milestones, a concern voiced by both practitioners and researchers \cite{Deshpande2019}. 

Uncovering relations between requirements primarily expressed in natural language presents hurdles due to the intricacies of human language, requiring manual effort to decipher connections. This task is time-consuming and error-prone, especially considering the cognitive load it poses in domain-specific, specialized environments~\cite{Tanjong2021,Frattini2020,Mokammel2018,Deshpande2020,Raatikainen2022}. Furthermore, relations come with distinct attributes like the semantics of the relation, their directionality and the degree of influence or strength, which might be difficult to infer from textual requirements. 
The role of Natural Language Processing (NLP) in this context is pivotal. By leveraging syntactic and semantic knowledge, NLP techniques hold the potential to decipher requirement relations embedded within natural language. Linguistic-based knowledge further empowers information extraction methods by enabling the automatic identification of these relations from vast amounts of textual data. Such techniques, ranging from rule-based approaches to advanced machine learning, deep learning models and, ultimately, Large Language Models (LLM), offer a spectrum of strategies to capture these relations effectively. Consequently, this chapter aims to shed light on these techniques, equipping researchers and practitioners in Requirements Engineering (RE) with valuable insights into NLP's capability to unravel the intricate web of requirement relations and, in doing so, enhance the quality of software development processes.

The rest of the chapter is structured as follows. Section \ref{sec:fundamentals} delves into the fundamentals of requirements relations. Section \ref{sec:nl-techniques} depicts the role of NLP knowledge generation techniques to support relation extraction. Section \ref{sec:methods} elaborates on the different NLP-based information extraction methods covered by state-of-the-art literature in the field. 
Section \ref{sec:challenges} enumerates the most relevant challenges from the state-of-the-art and future research directions. Finally, Section \ref{sec:conclusions} highlights the main conclusions from this chapter. 

\section{Fundamentals}
\label{sec:fundamentals}

\subsection{Requirements relations}
\label{sec:relations}

Typically, the knowledge and formalization of relations between requirements is mainly covered in detail as a specific type of \textit{traceability} between requirements. Specifically, the Handbook Requirements Management from the International Requirements Engineering Board (IREB)~\cite{IREB} refers to three types of requirement \textit{traces}: (i) back to its origins, (ii) forward to its implementation in design, code and associated tests, and (iii) to requirements it depends on~\cite{Glinz2022Glossary}. This \textit{traceability to other requirements} can refer to either the same or a different level of detail. 
Similarly, Pohl uses the term \textit{relationship} to refer to the interrelation among textual requirements, model-based requirements, and the interplay between textual and model-based requirements~\cite{Pohl2010}, making explicit the multiple dimensions of requirement relations at different levels of detail and specification~\cite{Buhne2022}. 

While \textit{relation} is the preferred term for referencing semantic-agnostic associations between requirements (i.e., with independence of the meaning of such relation), the term \textit{dependency} is broadly used in literature at different levels of detail. Mainly, \textit{dependency} is used to refer to a specific relation type expressing that a requirement should be fulfilled first because another requirement depends on it~\cite{Glinz2022}, or more generally, to a specific traceability link type~\cite{VanLamsweerde2009,Pohl2010,Buhne2022}. 
Moreover, relations between requirements extending the semantics of this prioritization-based perspective of the term dependency are broadly covered, including but not restricted to conflicting~\cite{Pohl2010,Robertson2012,Dick2017,Buhne2022} or duplicated~\cite{Dick2017,Buhne2022} requirements (we discuss relation types in more detail in Section \ref{sec:types}), among others.
Actually, researchers in the RE field generally use \textit{relation}~\cite{Abeba2022,Fischbach2021a,Schlutter2021,Guo2021,Cui2021,Jadallah2021,Schlutter2020,Sonbol2020,Frattini2020,Alhoshan2019,Alhoshan2018} and \textit{dependency}~\cite{Raatikainen2022,Ogawa2022,Tanjong2021,Fischbach2021,Deshpande2020,Asyrofi2020,Priyadi2019,Motger2019,Deshpande2019,Tahvili2018,Sree-Kumar2018} as indistinctive terms to refer to semantic associations between requirements, with independence of the covered relation types. Alternative terms like \textit{entailment}~\cite{Firmawan2022} or \textit{link}~\cite{Mokammel2018}, while also used, are less frequent. In this chapter, we will use the term \textit{relation} to refer to all types of textual requirements associations. Moreover, we will focus exclusively on relations between textual requirements and NLP-based strategies for their management.


\begin{defs*}{Relation}
A \textbf{relation} between a pair of requirements $r_i \xrightarrow[]{d} r_j$ is defined as a directed typed link from $r_i$ to $r_j$, where $d$ is the syntactic form of a specific type of semantic association between $r_i$ and $r_j$.
\end{defs*}

While requirement relations are defined as directed links, some specific relation types $d$ are defined as bidirectional, reciprocal relations, meaning that $r_i \xrightarrow[]{d} r_j$ implies that $r_j \xrightarrow[]{d} r_i$ for that same relation type $d$.


\subsection{Relation types}
\label{sec:types}

Building upon the foundations laid by the RE community, and using the definition in Section \ref{sec:relations}, we convey into the enumeration\footnote{This enumeration of requirement relations is not intended to be exhaustive, but representative of the most common relation types subject to be automatically detected by NLP-based techniques according to the scientific literature in the field.} of the following requirement relation types $d$, including whether they are unidirectional (U) or bidirectional (B) relations:

\begin{itemize}
    \item  \textbf{Requires} (U): the fulfilment of $r_i$ requires the fulfilment of $r_j$. This relation indicates that $r_j$ must be successfully addressed before $r_i$ can be considered complete. This is the most commonly addressed relation type~\cite{Pohl2010,VanLamsweerde2009,Glinz2022}, also referred to simply as \textit{depends}~\cite{Buhne2022,Glinz2022Glossary}.
    \item \textbf{Conflicts} (U): the fulfilment of $r_i$ restricts - without excluding - the fulfilment of $r_j$ ~\cite{Pohl2010,Robertson2012,Dick2017,Buhne2022}. This relation indicates the existence of limitations that need to be refined for resolving the conflicts between both requirements.
    \item \textbf{Contradicts} (B): $r_i$ and $r_j$ are mutually exclusive~\cite{Buhne2022} (i.e., $r_i$ \textit{conflicts} with $r_j$ to the limit of exclusion). This relation indicates that satisfying one requirement would lead to the violation of the other.
    \item \textbf{Is\_a\_variant} (B): $r_i$ serves as an alternative to $r_j$, providing a variant form to fulfil the same underlying need or purpose~\cite{Buhne2022}. This relation suggests that both requirements can be considered as an alternate option or choice in place of each other.
    \item \textbf{Is\_similar} (B): $r_i$ replicates - partially or totally - the content described by $r_j$~\cite{Dick2017,Pohl2010,Buhne2022}. This relation implies that $r_i$ conveys information or describes a desired behaviour also present in $r_j$, resulting in redundancy or content repetition.
    \item \textbf{Details} (U): $r_i$ extends or refines the information in $r_j$~\cite{Buhne2022}. This relation implies that $r_i$ adds more specific or detailed aspects to $r_j$, contributing to a deeper understanding of its behaviour, constraints or implementation.
\end{itemize}

Table \ref{tab:req-rel-ex} provides some illustrative examples for each requirement relation type $d$ listed above.

\begin{table}[ht]
\centering
\caption{Requirement relation examples for each relation type $d$. These examples have been specifically generated for academic and illustrative purposes, using generic, context-agnostic relation types.}
\begin{tabular}{@{}l@{\hspace{0.4cm}}p{4.8cm}@{\hspace{0.4cm}}p{4.8cm}@{}}
\toprule
\textbf{Type $d$}            & \textbf{Requirement $r_i$}                                                           & \textbf{Requirement $r_j$}                                                    \\ \midrule
Requires                     & When the ``Submit" button is clicked, the form data should be sent to the server for processing.     & The form user interface must include a ``Submit" button. \\
Conflicts                    & The software shall have a max. response time of 2 seconds for user requests. & The software shall be capable of processing extremely large datasets efficiently. \\
Contradicts                 & The software shall provide fault tolerance through real-time, synchronous data replication. & The software shall provide fault tolerance through offline periodic backups.     \\
Is\_a\_variant              & The software can use PostgreSQL as its database system.         & The software can use MySQL as its database system.       \\
Is\_similar                  & The software shall log all user interactions for auditing.       & The software shall record user activities for analysis.   \\
Details                      &  The project shall follow ISO 31000 guidelines for risk management.                       & The project shall incorporate risk management practices. \\ \bottomrule
\end{tabular}
\label{tab:req-rel-ex}
\end{table}


\subsection{NLP-based relation extraction}

In the context of NLP for RE~\cite{Zhao2021}, and based on the previous formalization, NLP-based relation extraction refers to the application of computational linguistic techniques and information extraction methods to \textbf{automatically identify and classify specific $r_i \xrightarrow[]{d} r_j$ instances between pairs of textual requirements $r_i,r_j \in R$}.
This set $R = \{r_1, r_2, ..., r_n\}$ is the corpus of documents (i.e., textual requirements) of size $n$, and each document $r_i \in R$ is an individual document from the corpus. These techniques and methods aim to automatically identify requirements relations $r_i \xrightarrow[]{d} r_j$ by leveraging NLP algorithms and methodologies, including syntactic parsing, semantic analysis, conceptual modelling and machine- and deep-learning models. 

In the rest of this chapter, we will discuss the most prominent strategies used for NLP-based relation extraction, which we structure into two sequential stages. 

\begin{enumerate}
    \item \textbf{NLP Knowledge Representation.} From the computational linguistics point of view, these techniques relate to the processing of textual documents to build a structured knowledge representation from a corpus $R$, based on the syntactic features and semantic knowledge from these textual representations. We categorize NLP-based techniques (Section \ref{sec:nl-techniques}) into syntactic (Section \ref{sec:syntactic}) and semantic (Section \ref{sec:semantic}) techniques, both of them acting after a pre-processing step (Section \ref{sec:preprocess}). 
    \item \textbf{Information Extraction.} These methods illustrate the practical application of information extraction methods using the structured syntactic- and semantic-based representation from textual requirements to identify and classify requirement relations. We categorize relation extraction methods (Section \ref{sec:methods}) into retrieval-based (Section \ref{sec:rule}) and machine-learning-based (Section \ref{sec:ml}) methods. 
\end{enumerate}

\begin{figure*}[htbp]
\centerline{\includegraphics[width=\textwidth]{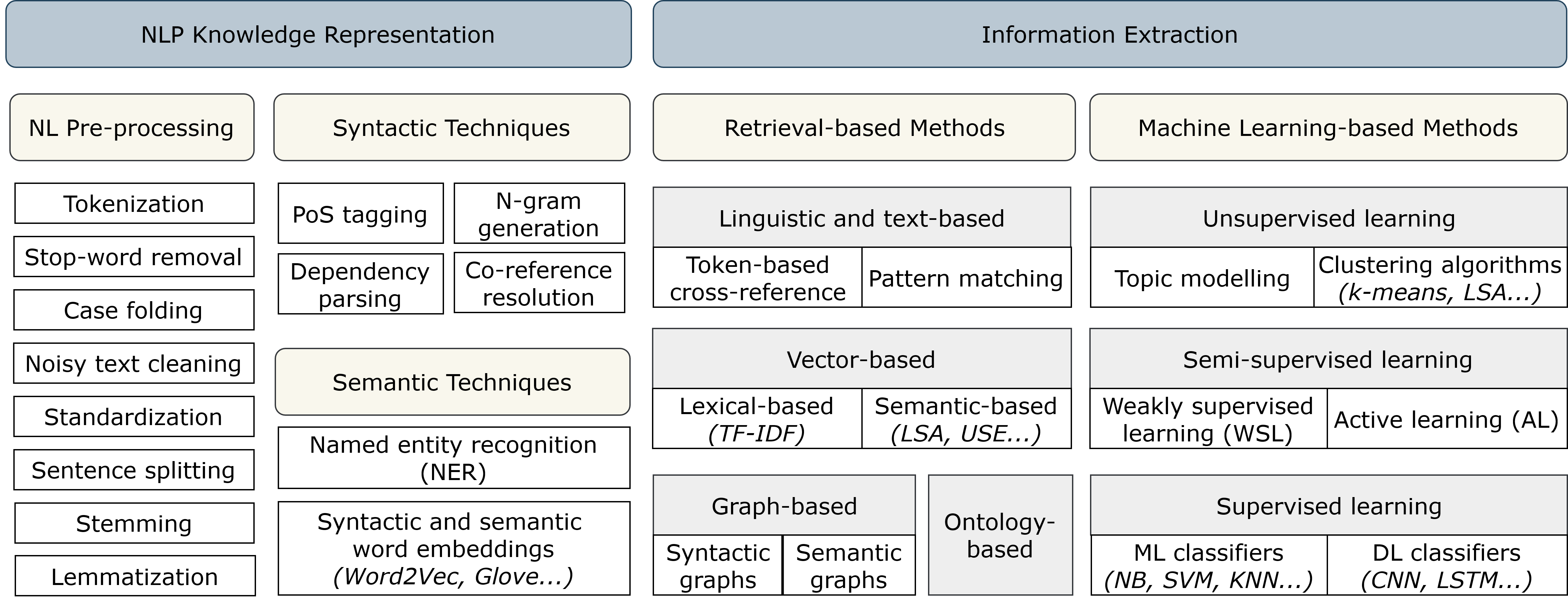}}
\caption{Summary of NLP techniques and relation extraction methods.}
\label{fig:rel-ex}
\end{figure*}

Figure \ref{fig:rel-ex} illustrates a summary of the NLP techniques and relation extraction methods defined in this chapter.

\subsection{Sample data set: PURE}
\label{sec:pure}

For illustrative purposes, we will use a sample selection of annotated requirement pairs from the PURE (PUblic REquirements) data set~\cite{Ferrari2017}, manually annotated by Deshpande et al.~\cite{Deshpande2019}. This sample contains a sub set of requirements from the functional requirements specification document (SRS) of the European Rail Traffic Management System (ERTMS/ETCS)~\cite{ErtmsETCS}. The data set is composed of two sub sets documenting 190 unique requirements: one for binary classification (i.e., non-related vs. related requirements) and one for multiclass classification (i.e., multiple requirement relation types). The binary data set contains 10,859 requirement relations (9,606 non-related, 1,253 related), while the multiclass data set contains 4,432 requirement relations (3,720 non-related, 378 \textit{require} relations, 334 \textit{similar} relations). Table \ref{tab:related-examples} contains some examples illustrating the different annotated requirement relation types in the data set. Complementary, we provide access to the data set using a Jupyter Notebook\footnote{Available at: \url{https://github.com/quim-motger/NLP4RE\_RelationExtraction}}. This notebook and data set will be used for exemplifying some of the techniques and methods depicted in this chapter for relation extraction.

\begin{table}[ht]
\centering
\caption{Requirement relation examples from the PURE annotated data set~\cite{Deshpande2019}.}
\begin{tabularx}{\textwidth}{p{0.1\textwidth}@{\hspace{8pt}}p{0.4\textwidth}@{\hspace{8pt}}p{0.4\textwidth}}
\toprule
\textbf{Type $d$} & \textbf{Requirement $r_i$} & \textbf{Requirement $r_j$} \\
\midrule
\textit{none} & The driver shall be able to select train data entry on the DMI. & On lines fitted with RBC the ETCS trainborne equipment shall be able to transmit the location of the entire train to the RBC. \\
requires & The ETCS on-board shall be capable of receiving information about pantograph and power supply from the trackside. &  The information regarding lowering and raising of the pantograph and opening/closing of the circuit breaker shall be provided separately and in combinations. \\
is\_similar & The current operational status shall be indicated to the driver on the DMI. & A special indication shall be shown on the DMI. \\
\bottomrule
\end{tabularx}
\label{tab:related-examples}
\end{table}

\begin{tcolorbox}
[width=\linewidth, sharp corners=all, colback=white!95!black, boxrule=0.2pt, boxsep=1pt,left=2pt,right=2pt,top=2pt,bottom=2pt]
\includegraphics[height=10pt]{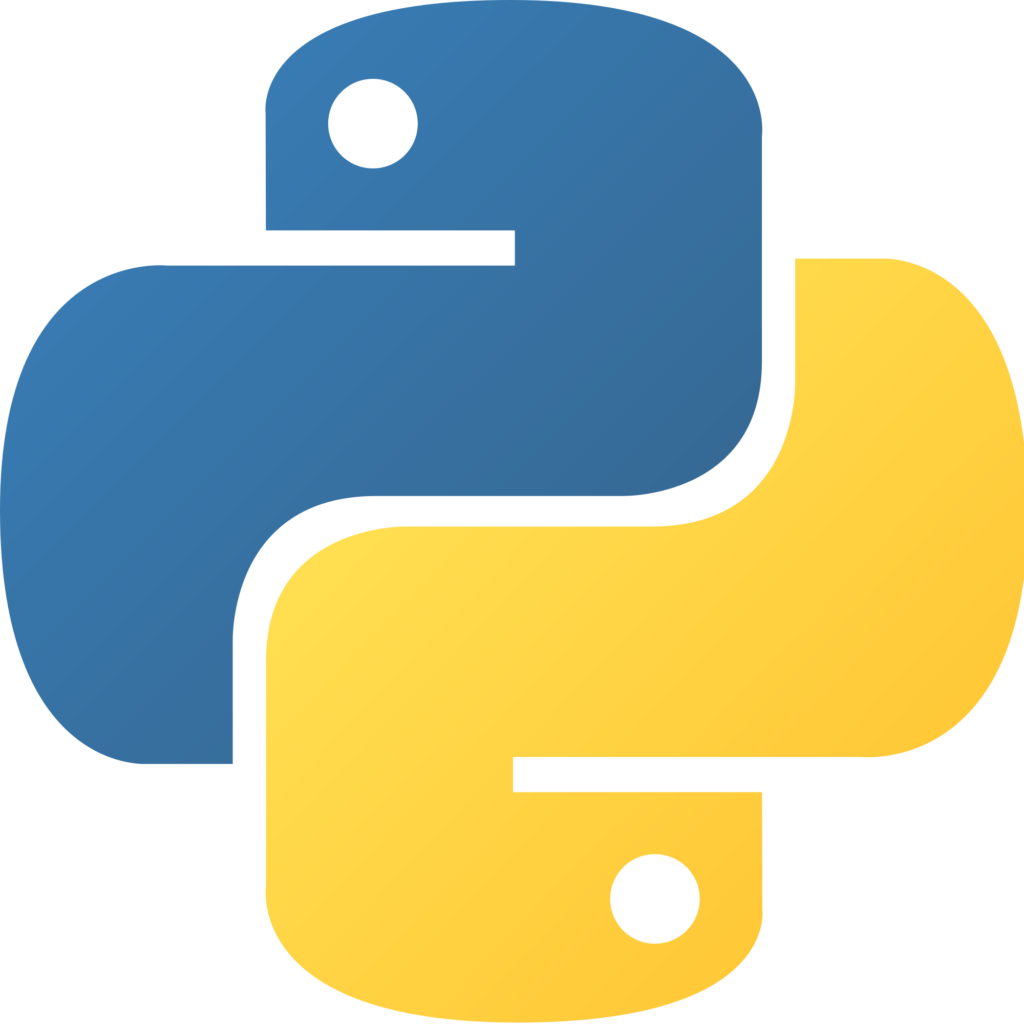}
Read and execute \textit{Section 2. Dataset} from the Jupyter Notebook to load and explore the PURE binary and multiclass datasets.
\end{tcolorbox} 
\section{NLP Knowledge Representation Techniques}
\label{sec:nl-techniques}

\subsection{Text Pre-processing Techniques}
\label{sec:preprocess}

Initial steps in most NLP tasks aim to prepare the text for further analysis and extraction of meaningful information. These include, but are not restricted to, the following tasks: tokenization, stop-word removal, case folding, noisy text cleaning, standardization, sentence splitting, stemming and lemmatization. 
In this stage, each requirement document from a given corpus $r \in R$ is processed through a NLP pipeline to generate a tokenized representation $T(r) = [t_1, t_2, ..., t_m]$, where each $t_i$ represents a token at the position $i$ in the original $r$.
Selection, combination and order of the aforementioned techniques is diversely represented in practical applications of relation extraction strategies. Solutions range from simple tokenization, either with traditional approaches like whitespace, punctuation-based tokenizers~\cite{Mokammel2018,Sree-Kumar2018} or LLM-based tokenizers~\cite{Fischbach2021}, to a full combination of these techniques~\cite{Abeba2022,Deshpande2020}. For effective relation extraction, it is crucial to strike a balance between preserving data integrity and removing noise, in order to maximize precision without sacrificing valuable information. A conservative approach might risk diminishing domain-specific nuances and human-input particularities that contribute to understanding context and identifying patterns, while a permissive strategy could overcomplicate the task by processing excessive noise and non-relevant text input.
This delicate balance underscores the importance of employing pre-processing techniques tailored to the data's nature, text quality, and specific requirements, aiming to optimize knowledge extraction accuracy while minimizing information loss or distortion.

For more details on text pre-processing techniques, we refer to Chapter 1 of this book.

\subsection{Syntactic-Based NLP Knowledge Representation Techniques}
\label{sec:syntactic}


\begin{defs*}{Syntactic-based techniques}
\textbf{Syntactic-based} techniques refer to the processing and utilization of syntactic and linguistic information to infer relations between requirements based on the structural aspects and rules governing the arrangement of words and phrases in a sentence or text. These techniques leverage syntactic patterns, rules and structures to serve as a foundation for understanding the structural and grammatical aspects of requirement relations.
\end{defs*}

Syntactic techniques focus on the extension of a tokenized representation of a requirement $r$, $T(r) = [t_1, t_2, ..., t_m]$, with syntactic feature annotations for each individual token $t_i$ and its syntactic association with other tokens $t_j, i \neq j$. One of the most common syntactic analysis practices in relation extraction is the combination of Part-of-Speech (PoS) tagging and dependency parsing. PoS tagging assigns a grammatical label $g_i$ to each $t_i$, indicating the syntactic category of the given token. Using this extended annotation, dependency parsing is formalized through a dependency tree $DT(r)$, which is defined as a rooted directed tree composed of token dependencies $DP(t_i, t_j, dep)$, where $dep$ is the type of dependency from $t_i$ to $t_j$. Again, we refer to Chapter 1 for more details.

\begin{figure*}[htbp]
\centerline{\includegraphics[width=\textwidth]{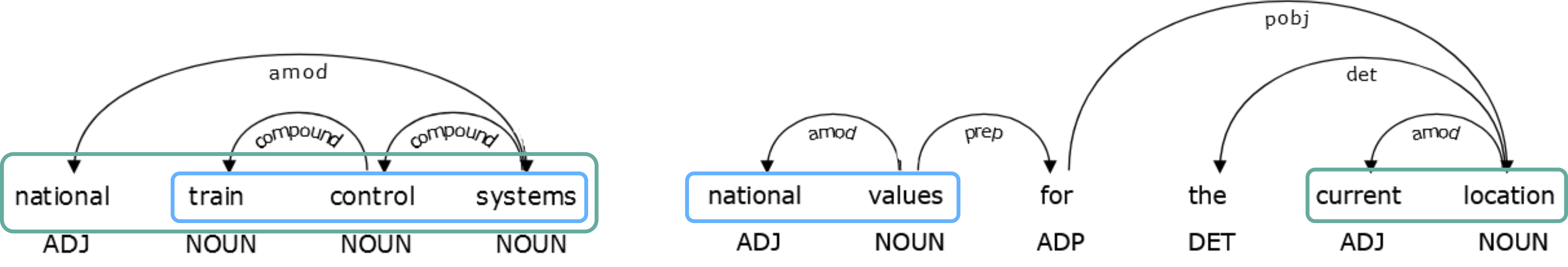}}
\caption{Examples of the generation of n-grams. On the left, a parent node (\textit{systems}) is the root of two nested n-grams composed by subsequent (\textit{control} and \textit{train}) and parallel (\textit{national}) direct term children. On the right, a non-relevant child (\textit{for}) from the root term (\textit{location}) associates the second n-gram (\textit{current} and \textit{location}) with the first one (\textit{national} and \textit{values}).}
\label{fig:n-gram}
\end{figure*}

Grammatical associations at sentence level using PoS tags are the most common syntactic-based features used in requirements relations~\cite{Ogawa2022,Fischbach2021,Schlutter2021,Guo2021,Cui2021,Schlutter2020,Sonbol2020,Asyrofi2020,Priyadi2019,Motger2019,Deshpande2019,Alhoshan2019,Sree-Kumar2018,Mokammel2018}. 
While syntactic annotations are mostly used to generate structured knowledge used by syntactic-based relation extraction methods (see Section \ref{sec:rule}), some techniques can also be used in isolation as relation extraction methods. These syntactic-based techniques include but are not limited to the following: 

\begin{itemize}
    \item \textbf{N-gram generation}: generating sequences of $m$ contiguous tokens $t_{i\rightarrow i+n-1}$ to capture contextual information and encapsulate more general ideas beyond individual tokens~\cite{Motger2019}. N-grams are typically generated based on pre-defined syntactic rules, whether explicitly set by generic and/or domain-specific PoS patterns (e.g., NOUN-compound-NOUN, NOUN-amod-ADJ) or implicitly inferred by term-based matching using pattern matching templates or ontologies (see Section \ref{sec:rule}). A special case of syntactic-based n-gram generation is \textbf{noun phrase chunking}, grouping tokens to form meaningful, compound noun phrases~\cite{Sree-Kumar2018,Mokammel2018}, encompassing pronouns (PRON), proper nouns (PROPN), or nouns (NOUN), which can be found alongside other tokens serving as modifiers, such as adjectives (ADJ) or additional nouns. Figure \ref{fig:n-gram} illustrates a couple of examples using n-gram generation, reflecting on the extraction of complex, compound entities and its hierarchy, which can later be used to refine the matching accuracy of related requirements beyond single-term, context-agnostic matches.
\end{itemize}

\begin{figure*}[htbp]
\centerline{\includegraphics[width=\textwidth]{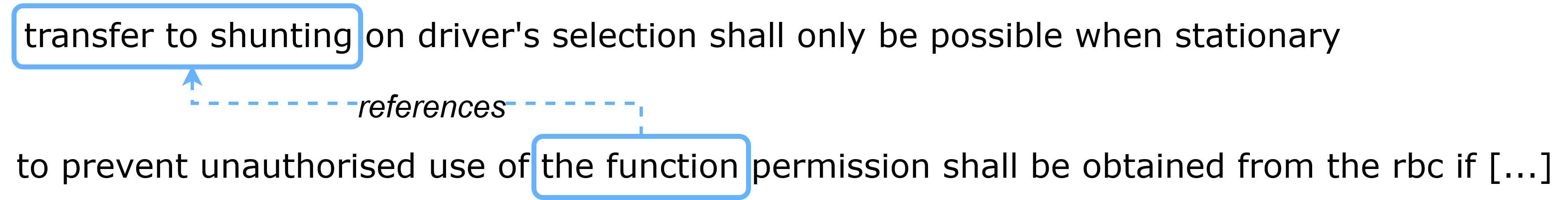}}
\caption{Example of cross-document coreference. In the second requirement, \textit{the function} refers to the core functionality \textit{transfer to shunting} mentioned in the first requirement, which is the immediate document predecessor.}
\label{fig:coref}
\end{figure*}

\begin{itemize}
    \item \textbf{Coreference resolution}: resolving references through pronouns or noun phrases to the same entity in a given requirement or between multiple requirements~\cite{Schlutter2021,Schlutter2020,Sonbol2020,Fritz2020}. While coreference resolution is a traditional NLP task for disambiguation, in the context of relation extraction it supports potential identification of concurrent references to the same entities between different requirement documents, known as cross-document coreference. The most popular strategy is the use of pre-trained neural models~\cite{NeuralCoref} for generic, syntactic-based coreferences~\cite{Sonbol2020}, extended with domain-specific rule-based techniques like context-based synonymy rules using mapping tables~\cite{Abbas2023} or location-based references using order and hierarchy between requirement documents~\cite{Ogawa2022}. In this case, syntactic coreferences do not allow for semantic analysis for inference of relation types. Figure \ref{fig:coref} illustrates a cross-document reference example through noun phrase disambiguation.
\end{itemize}

Generally, these techniques lay the groundwork for pattern-matching based techniques in which grammatical structures infer specific templates to express a potential relation (see Section \ref{sec:rule}). Nevertheless, they can also be used to generate enriched, relation specific word embeddings 
for either unsupervised~\cite{Cui2021} or supervised~\cite{Sree-Kumar2018} machine- and deep-learning approaches (see Section \ref{sec:ml}). 
In comparison with traditional purely lexical embeddings like TF-IDF (see Section \ref{sec:methods}), syntactic embeddings delve deeper into the relation between words based on the nuances of the language syntax.

\begin{tcolorbox}
[width=\linewidth, sharp corners=all, colback=white!95!black, boxrule=0.2pt, boxsep=1pt,left=2pt,right=2pt,top=2pt,bottom=2pt]
\includegraphics[height=10pt]{figures/colab.png}
Read and execute \textit{Section 3. Syntactic NLP Techniques} from the Jupyter Notebook to analyse the pre-processing and syntactic analysis of a sub set of requirement pairs from the PURE annotated data set using the English RoBERTa-based transformer pipeline from Spacy~\cite{Spacy}.
\end{tcolorbox} 

\subsection{Semantic-Based NLP Knowledge Representation Techniques}
\label{sec:semantic}


\begin{defs*}{Semantic-based techniques}
\textbf{Semantic-based} techniques involve the application of linguistic and knowledge-based approaches to infer requirement relations based on the meaning and interpretation of words, phrases, and sentences in a language. These techniques aim to capture the conceptual relationships, associations, and nuances of language, enhancing the comprehension and analysis of requirement relations by leveraging their inherent semantic knowledge.
\end{defs*}

\begin{figure*}[htbp]
\centerline{\includegraphics[width=\textwidth]{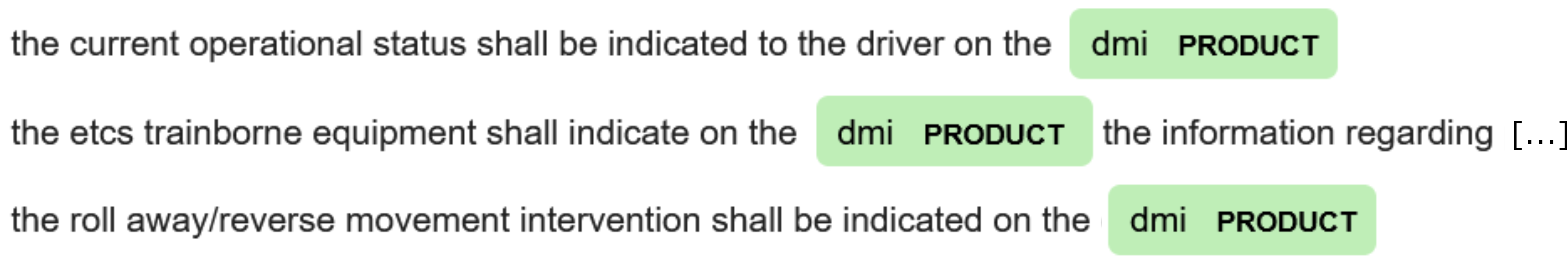}}
\caption{Example of named-entity recognition (NER). All reported requirements refer to the display of information in the Driver Machine Interface (DMI), an interface component between the driver and the ERTMS/ETCS system.}
\label{fig:ner}
\end{figure*}

Semantic techniques focus on the extension of $T(r) = [t_1, t_2, ..., t_m]$ with semantic feature annotations for each individual token $t_i$, as well as subsets of multiple tokens (i.e., n-gram combinations). Most common semantic-based techniques include the following:

\begin{itemize}
    \item \textbf{Named Entity Recognition (NER)}. At token level, a common example of semantic feature annotations are NER tags, which allow the identification of a set of tokens $T_{NER}(r) \in T(r)$ as named entities. These named entities include stakeholders, services, products, systems and software components, among others~\cite{Sree-Kumar2018}. They provide semantic knowledge about the subject and/or object of a relation, but not about the semantics of the relation type. NER can be performed using pre-trained models with generic tags (e.g., organization, product, quantities, cardinals) or fine-tuned models for domain-specific terms and/or labels. For example, in the context of ERTMS/ETCS~\cite{ErtmsETCS}, references to systems or components like LZB (Linienzugbeeinflussung), RBC (Radio Block Center) or DMI (Driver Machine Interface) might not be detected without an extension of pre-trained NER models. Figure \ref{fig:ner} illustrates potential related requirements through common NER assigned tags.
\end{itemize}

\begin{itemize}
    \item \textbf{Syntactic and semantic word embeddings}. 
    Embedding generation from pre-trained models like Word2Vec~\cite{Tanjong2021,Sonbol2020,Alhoshan2018} or GloVe~\cite{Jadallah2021} map words with similar meanings or contexts to similar vector representations, enabling the capture of semantic relationships. These vector-based representations can later be used for multiple syntactic and semantic knowledge inference tasks, including semantic similarity~\cite{Schlutter2020,Priyadi2019,Motger2019,Alhoshan2019} using cosine, Euclidean or Manhattan similarity algorithms~\cite{Alhoshan2019}. Semantic similarity might not only be used to identify potential duplicate relations $r_i \xrightarrow[]{similar} r_j$ at requirement level, but also to automatically identify individual tokens $t_i$ or n-grams $t_{i\rightarrow i+n-1}$ from $r_i$ that are semantically similar to another subset of tokens from another requirement $r_j$, meaning that these probably refer to highly-related or even equivalent entities. More recent solutions based on the use of LLM use the embedding generation suited for the selected LLM (see Section \ref{sec:ml}). 
\end{itemize}

\begin{tcolorbox}
[width=\linewidth, sharp corners=all, colback=white!95!black, boxrule=0.2pt, boxsep=1pt,left=2pt,right=2pt,top=2pt,bottom=2pt]
\includegraphics[height=10pt]{figures/colab.png}
Read and execute \textit{Section 4. Semantic NLP Techniques} from the Jupyter Notebook to explore semantic analysis (i.e., NER, semantic similarity) of a sub set of requirement pairs from the PURE annotated data set using Word2Vec.
\end{tcolorbox} 

In conclusion, syntactic techniques analyse the structural aspects of language, unveiling relationships embedded in grammatical patterns. Conversely, semantic techniques delve deeper into word meanings, capturing the underlying nuances of language in domain-specific scenarios. Both syntactic and semantic techniques can be used either in isolation or as hybrid approaches, combining the acquired knowledge from both perspectives. Deciding on the use of these techniques will be conditioned by the relation extraction methods applied in each context (see Section \ref{sec:methods}). 
\section{Information Extraction Methods}
\label{sec:methods}

From an information extraction standpoint, relation extraction methods can be majorly classified into two core categories: retrieval-based (Section \ref{sec:rule}) and machine-learning-based (Section \ref{sec:ml}) methods. 

\subsection{Retrieval-Based Information Extraction Methods}
\label{sec:rule}


\begin{defs*}{Retrieval-based methods}
\textbf{Retrieval-based} methods involve the application of predefined rules, patterns, ontologies or linguistic structures to analyse and extract potential relations from a corpus $R$. 
\end{defs*}

\subsubsection{Linguistic and Text-based Methods.} These deterministic approaches leverage syntactic and semantic extended features from each requirement $r$, such as the tokenization $T(r)$, the dependency tree $DT(r)$, and the set of named entities $T_{NER}(r)$. The objective is to identify coreferent or semantically related statements that imply a relation between requirement pairs $r_i, r_j \in R$. As illustrated in Figure \ref{fig:rel-ex}, common methods in this category include:

\begin{itemize}
    \item \textbf{Token-based cross-reference detection.} This formalization is based on the detection of tokens indicating a crossed reference between $r_i$ and $r_j$. The most common example of cross-reference implies that a sub set of tokens $t \subseteq T(r_i)$ refers to the same sub set of tokens $t \subseteq T(r_j)$. This illustrates cross-document references to agents or components affecting or affected by related requirements (e.g., DMI in Figure \ref{fig:ner}). 
    Additionally, a crossed reference might also imply that a sub set of tokens $t \subseteq T(r_i)$ refers to a specific metadata field (e.g., the ID) from $r_j$. This refers to implicit relations involving explicit references to metadata textual properties, including a requirement identifier, category, product or system to which the requirement belongs to.
    \item \textbf{Pattern matching.} This method involves utilizing multiple text-based NLP-based properties, including the resulted tokenization $T(r_i)$ or dependency tree $DT(r_i)$, to apply automatic matching with a set of pre-defined, context-specific rules compliant with the formalization of a relation within a given domain. The most common pattern matching technique is keyword matching, where a set of domain-specific pre-defined words or tokens are used as a dictionary of potential relevant keywords for relation formalization~\cite{Ogawa2022,Guo2021,Priyadi2019,Sree-Kumar2018}. These words can either relate to co-location of entities and requirement artefacts with respect to the original requirement (e.g., ``\textit{below}'', ``\textit{the following}'', ``\textit{in the document above}''), or with respect to another document (e.g., ``\textit{appendix}'', ``\textit{attached file}'')~\cite{Ogawa2022}. Complementarily, information in $DT(r_i)$ can be used to extend keyword matching by filtering out non-relevant appearances of specific keywords (e.g., by excluding matches for which the keyword is not the ROOT element of the sentence in order to exclude non-principal clauses). 
\end{itemize}

\begin{figure*}[htbp]
\centerline{\includegraphics[width=\textwidth]{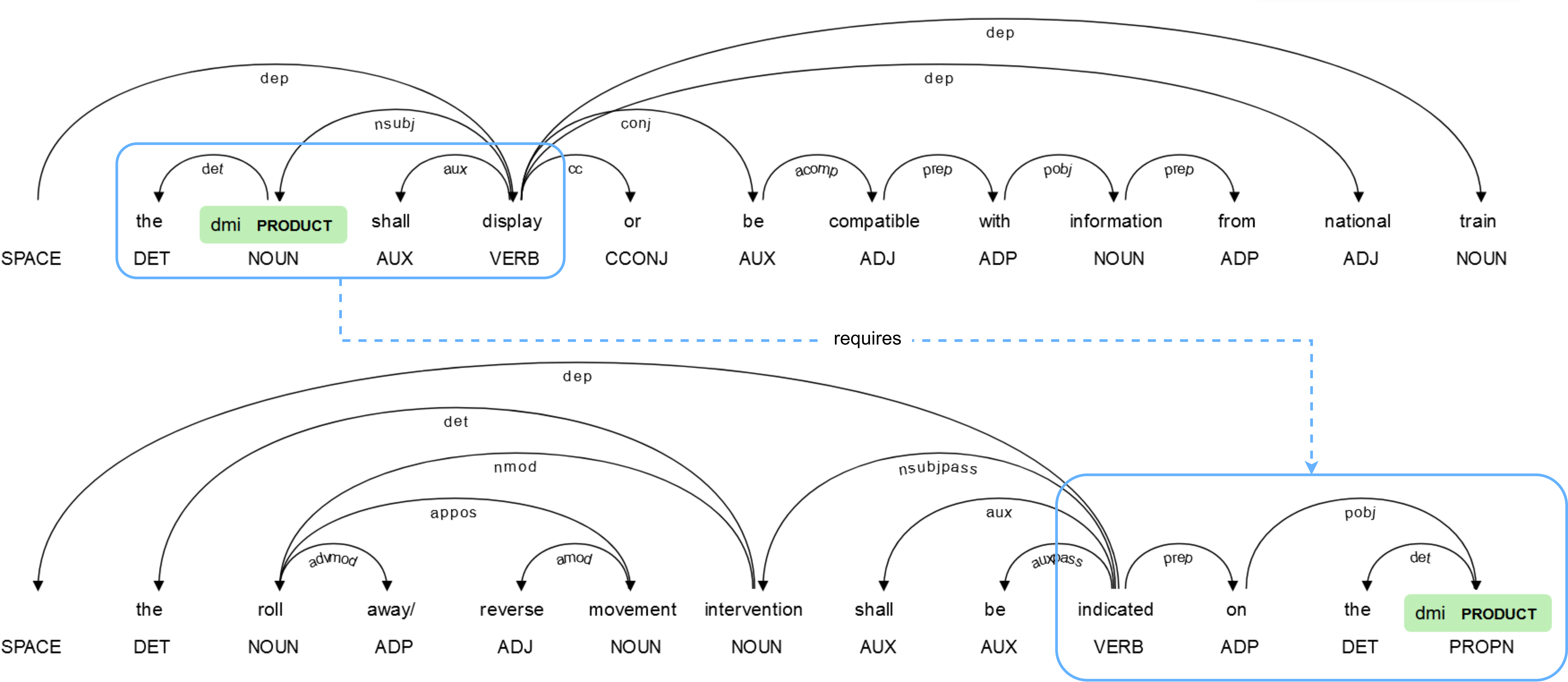}}
\caption{Dependency tree and named entity recognition results for a pair of annotated requirements $r_i \xrightarrow[]{requires} r_j$ from the PURE data set.
}
\label{fig:sync-sem-req}
\end{figure*}

Figure \ref{fig:sync-sem-req} illustrates a syntactic (dependency tree) and semantic (NER) hybrid combination of pattern-matching knowledge inference for relation detection of a $requires$ relation between a pair of requirements $r_i, r_j \in R$ from the annotated PURE data set. NER task allows for the detection of relevant context-dependent named entities $T_{NER}(r_i) \subseteq T(r_i)$ and $T_{NER}(r_j) \subseteq T(r_j)$ (i.e., the \textit{Driver Machine Interface} or DMI) without extended pre-training or fine-tuning. A pair of requirements for which $T_{NER}(r_i) \cap T_{NER}(r_j) \neq \O$ is reporting two requirement documents for which an explicit reference to the same component is being reported. For a more accurate keyword detection, a more fine-grained categorization of entity types, or for a more specialized domain, supervised training on a NER pipeline could potentially improve component identification.

Intersection on named entities $T_{NER}(r_i) \cap T_{NER}(r_j)$ can be used as a form of in-text explicit cross-reference. Nevertheless, syntactic knowledge can either be used as a complement or a stand-alone approach to identify linguistic patterns in which a relation to another requirement part or document might be suggested. Given a named entity $t_{ner} \subseteq T(r_i)$, for all $DP(t_j, t_k, dep)$ where $j \neq k$ and $t_j \in t_{ner}$ or $t_k \in t_{ner}$, a sub-set of pre-defined dependency patterns can be used to determine occurrences of $t_{ner}$ syntactically relevant for the materialization of a relation. While the effectiveness of distinct patterns is typically context dependent (e.g., authorship of requirements, formalization style, extension, typology of documentation), general rules include lexical intersection on the root element and/or direct syntactic dependencies, and overlapping of common, shared syntactic patterns between requirements.
In Figure \ref{fig:sync-sem-req}, a specific named entity (i.e., the DMI) acts as nominal subject ($nsubj$) of the ROOT verb element as displayed in $DT(r_i)$ (i.e., \textit{``display"}), while that same named entity, by transitivity, acts as the object of the ROOT verb element in $DT(r_j)$ (i.e., \textit{``indicated"}). 

\begin{tcolorbox}
[width=\linewidth, sharp corners=all, colback=white!95!black, boxrule=0.2pt, boxsep=1pt,left=2pt,right=2pt,top=2pt,bottom=2pt]
\includegraphics[height=10pt]{figures/colab.png}
Read and execute \textit{Section 5.1. Linguistic and Text-based Methods} from the Jupyter Notebook for practical application of a NER-based cross-reference detection using pattern matching to identify related requirements.
\end{tcolorbox} 

\subsubsection{Vectorization Methods.} These methods involve creating vector-based representations of a requirement corpus $R$ using the tokenized, pre-processed representation of all $r_i \in R$ as input. These representations are then used to build lexical- and/or semantic-based vector space models of the requirement documents.

\begin{itemize}
    \item \textbf{Lexical-based detection.} This method involves applying statistical techniques to identify relations between requirements. 
    One of the most common approaches is the TF-IDF (Term Frequency-Inverse Document Frequency) vector space model. The TF-IDF score for a term $t_j \in T(r_i)$ is calculated as follows:

    \begin{equation}
        \text{\textit{TF-IDF}}(t_j, r_i) = \text{\textit{TF}}(t_j, r_i) \cdot \text{\textit{IDF}}(t_j)
    \end{equation}

    where $TF(t_j, r_i)$ is the term frequency of term $t_j$ (i.e., the number of occurrences of term $t_j$ in requirement $r_i$), and IDF($t_j$) is the inverse document frequency of term $t_j$ across the set of requirements, defined as:
    
    \begin{equation}
        \text{\textit{IDF}}_{t_j} = log_2(\frac{n}{n_j})
    \end{equation}
    
    The TF-IDF score can then be used to identify relationships between requirements by looking for words that have high TF-IDF scores (i.e., passing a context-dependent threshold score $thr$) in both requirements.
    \item \textbf{Semantic-based detection.} This method involves using semantic-based techniques to generate requirements embeddings. This can be done through general purpose, unsupervised approaches like Latent Semantic Analysis (LSA) and Universal Sentence Encoder (USE), or by employing pre-trained models such as Word2Vec, GloVe or FastText. Numerical semantic vector representations can either be used to identify relations between requirements by applying statistical measures (e.g., using similarity measures) or as input to machine-and deep-learning based methods (see Section \ref{sec:ml}). 
\end{itemize}

\begin{tcolorbox}
[width=\linewidth, sharp corners=all, colback=white!95!black, boxrule=0.2pt, boxsep=1pt,left=2pt,right=2pt,top=2pt,bottom=2pt]
\includegraphics[height=10pt]{figures/colab.png}
Read and execute \textit{Section 5.2. Vectorization Methods} from the Jupyter Notebook to analyse a Word2Vec based illustration of related requirements in a vector-space representation.
\end{tcolorbox} 

\subsubsection{Graph-based Methods.} These methods leverage the structural relationships and dependencies within the requirement corpus $R$ to identify and extract meaningful connections between requirements. Two main approaches are considered:

\begin{itemize}
    \item \textbf{Syntactic graphs.} This method involves creating a weighted, directed graph model using the entire requirement corpus $R$ and the dependency tree for each requirement $DT(r_i)$ for $r_i \in R$. In syntactic or PoS-based graphs, nodes are tokens $t$ and edges are the syntactic dependencies between each pair of tokens $DP(t_i, t_j, dep)$, which might be weighted using frequency-related criteria (e.g. logarithmic occurrence frequency of each edge in $DT(r_i)$~\cite{Cui2021}). This method can either be used in isolation as an approach to identify dependencies using pattern-matching techniques~\cite{Asyrofi2020}, or as an input to generate relation-specific, syntax-enhanced embeddings of requirements for machine- and deep-learning based strategies~\cite{Cui2021} (see Section \ref{sec:ml}). The combination of graph-based requirements modelling with pattern matching techniques (e.g., keyword matching, cross-reference detection) allows for efficient updates and automated detection of potential dependencies based on the insertion of a new edge whose target node is a domain-specific relevant word. 
    \item \textbf{Semantic graphs.} This method involves identifying and extracting relationships between entities or concepts in text using semantic role labelling (SRL) and semantic relation graphs. SRL identifies semantically relevant predicates and associates their arguments with specific roles within a sentence. These roles provide information about the nature of the relationship between the predicate and its arguments. In the context of relation extraction, SRL helps in identifying the key entities involved in a relationship and their respective roles. The semantic relation graph represents the extracted relationships in the form of a directed graph, where predicates (verbs) are represented as nodes and their associated arguments (e.g., noun and person phrases) as additional, connected nodes. The graph also captures coreference relations and deduplicates arguments. 
\end{itemize}

\begin{figure*}[htbp]
\centerline{\includegraphics[width=\textwidth]{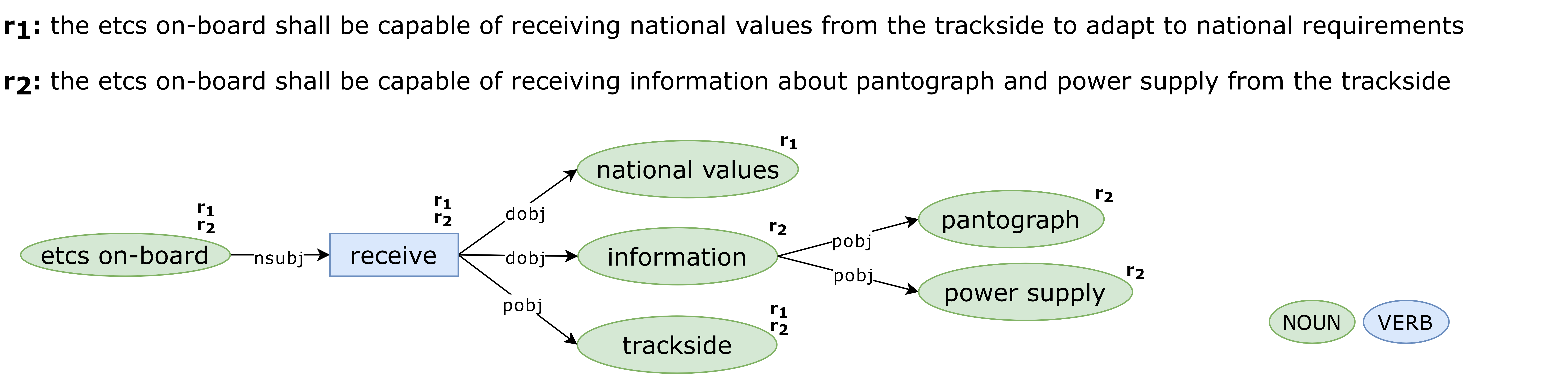}}
\caption{Example of a relational graph instance for a $r_i \xrightarrow[]{requires} r_j$ relation. The example illustrates the node density caused by shared entities among requirements (i.e., subject \textit{ETCS on-board}, verb \textit{receive}, object \textit{trackside}). 
}
\label{fig:graph}
\end{figure*}

Figure \ref{fig:graph} illustrates a reduced graph-based instance. 
This structure enables the representation and analysis of complex relationships and supports graph-based algorithms for further processing focused on semantic search, like spreading activation~\cite{Schlutter2021,Schlutter2020}. 
By initiating activation at specific nodes in the graph, such as predicates or key entities, the activation spreads along the edges, influencing and activating neighbouring nodes. This activation can be used to measure the strength or relevance of connections between entities, identify related concepts, or retrieve additional information from the graph.

\begin{tcolorbox}
[width=\linewidth, sharp corners=all, colback=white!95!black, boxrule=0.2pt, boxsep=1pt,left=2pt,right=2pt,top=2pt,bottom=2pt]
\includegraphics[height=10pt]{figures/colab.png}
Read and execute \textit{Section 5.3. Graph-based Methods} from the Jupyter Notebook for practical application of the construction of a SRL graph to identify related requirements.
\end{tcolorbox} 

\subsubsection{Ontology-based Methods.} 
\begin{figure*}[htbp]
\centerline{\includegraphics[width=\textwidth]{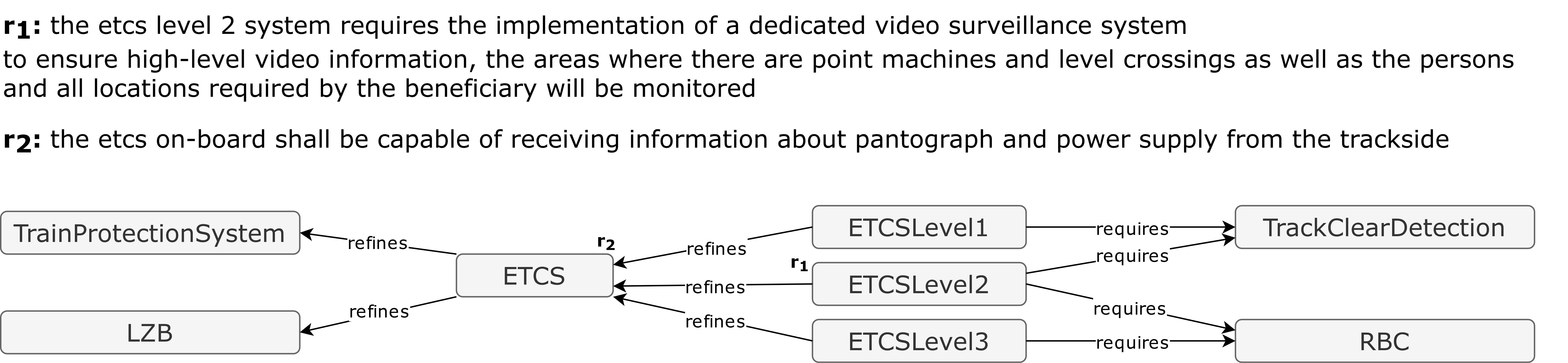}}
\caption{Example of a $r_i \xrightarrow[]{refines} r_j$ relation detected using an ontology for the ETCS/ERTMS domain. The \textit{ETCSLevel2} reference from $r_i$ refines implementation details from the information received referenced by $r_j$.}
\label{fig:ontology}
\end{figure*}

These methods use pre-defined ontologies as structured representations of domain-specific knowledge used as gold-standards for relation extraction. These techniques leverage both semantic relationships and hierarchical structure defined in ontologies to infer and categorize relations among requirements concepts. Generic lexical databases like WordNet~\cite{WordNet} can be used to infer linguistic relations such as synonymy (e.g., ``\textit{similar}"), hypernymy (e.g., ``\textit{details}") or hyponymy (e.g., ``\textit{variant}", ``\textit{conflicts}")~\cite{Shah2020}. On the other hand, domain-specific ontologies overcome the limitations of generic knowledge bases by modelling semantically-oriented relations between domain-specific terms, including agents, products, systems or components, among others~\cite{Deshpande2020,Motger2019,Sree-Kumar2018}. Design of domain-specific ontologies is typically done using the Web Ontology Language (OWL)~\cite{OWL}, a semantic web language specifically designed for representing and encoding knowledge in a machine-readable format. OWL is based on a formal logical foundation of expressive features, including various classes, properties, and relationships, and is built upon the Resource Description Framework (RDF)~\cite{RDF}. 
Ontology approaches use conceptual clustering and similarity analysis to match each pre-processed requirement $r_i \in R$ with the concepts of the ontology, including multiple n-gram combinations~\cite{Motger2019}.
Figure \ref{fig:ontology} illustrates a relation detection using an ontology example for the PURE dataset domain. 
In comparison with syntactic and semantic graph approaches, ontologies allow for effective domain-specific modelling to provide structured knowledge on specific industrial and scientific environments. Nevertheless, designing, generating and maintaining a domain-specific ontology is time-consuming and requires the involvement of domain experts.

\begin{tcolorbox}
[width=\linewidth, sharp corners=all, colback=white!95!black, boxrule=0.2pt, boxsep=1pt,left=2pt,right=2pt,top=2pt,bottom=2pt]
\includegraphics[height=10pt]{figures/colab.png}
Read and execute \textit{Section 5.4. Ontology-based Methods} from the Jupyter Notebook for practical application of an ontology-based semantic matching to identify related requirements.
\end{tcolorbox} 






\subsection{Machine Learning-Based Information Extraction Methods}
\label{sec:ml}


\begin{defs*}{Machine- and deep-learning-based methods}
\textbf{Machine- and deep-learning-based} methods encompass approaches that utilize machine- and deep-learning models to automatically learn and predict potential requirements relations. These methods leverage supervised, unsupervised, or weakly supervised learning approaches to detect and classify requirement relations within a given corpus. 
\end{defs*}

\subsubsection{Unsupervised Methods.} 
\begin{figure*}[htbp]
\centerline{\includegraphics[width=\textwidth]{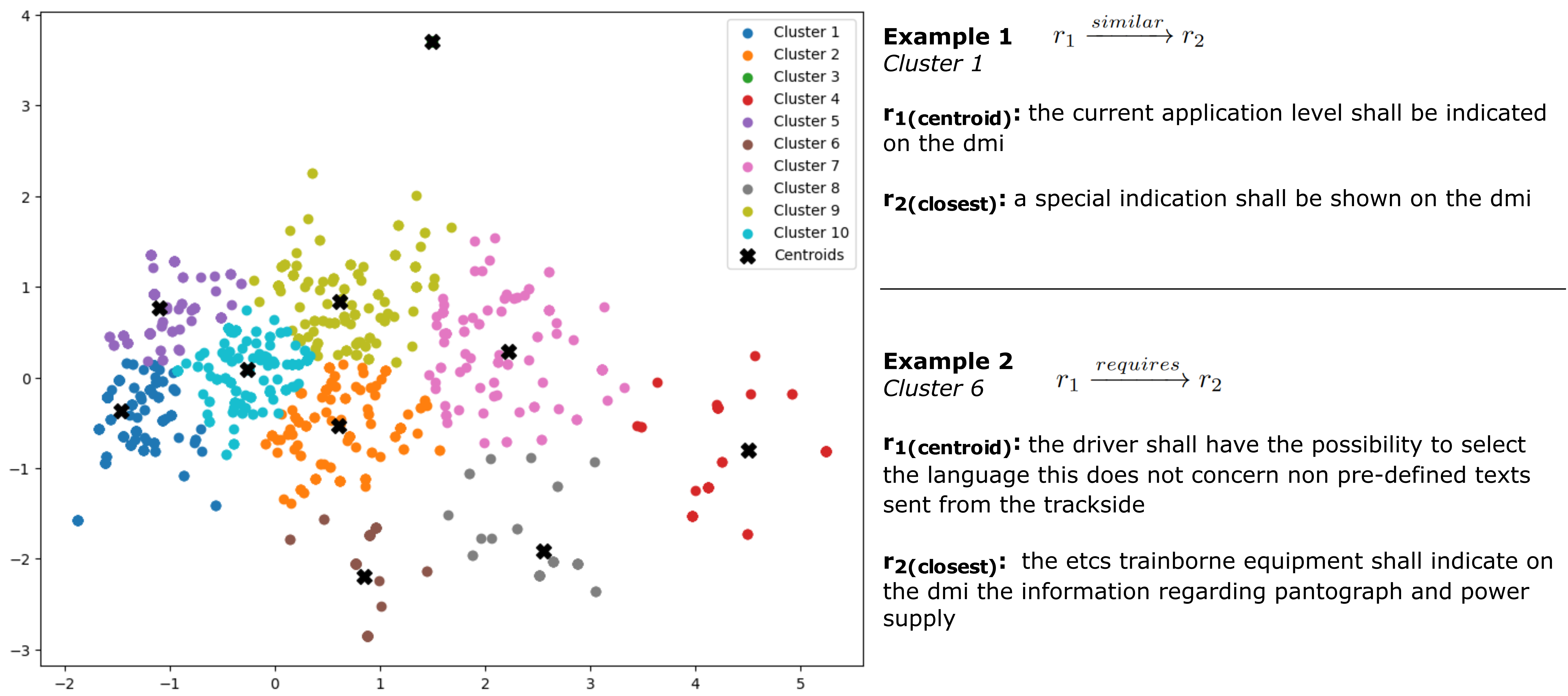}}
\caption{Output of k-means clustering (k=10) for a reduced sample set of 117 requirements from the PURE data set. Some examples for which an annotated relation exists have been highlighted.}
\label{fig:kmeans}
\end{figure*}

Unsupervised methods for relation extraction aim to discover patterns and relationships within the data without the use of labelled pairs of related requirements. These methods often involve clustering, topic modelling, or graph-based algorithms to identify and group similar instances based on their inherent similarities. As shown in Figure \ref{fig:rel-ex}, unsupervised methods include:

\begin{itemize}
    \item \textbf{Topic modelling.} Enriched syntax-enhanced and relation-specific embeddings from $R$ can be used as input for topic modelling algorithms~\cite{Cui2021}. These embeddings can be generated using enriched graph structures (see Section \ref{sec:semantic}) as input to deep learning models like Graph Convolutional Networks (GCN) using as training objective the target token for a given token $t \in T(r_i)$. These embeddings can later be used as input for modified multi-head self-attention layers to generate representations of requirement sentences and relations, employing block division based on entity pairs and incorporating self-attention mechanisms, max pooling, and linear transformations to encode and interact between sentences and relations for inference. 
    Notice that this technique is agnostic to the semantics of the relation $d$, meaning that reported relations are not categorized into a particular type.
    \item \textbf{Clustering algorithms.} Vectorization techniques can also serve as a contribution for enriched clustering algorithms. For instance, LSA enables clustering requirements into groups of similar requirements based on their semantic concepts~\cite{Mokammel2018}. Complementarily, a clustering algorithm like k-means can be applied using the LSA-generated vectors as input. The algorithm identifies the centroids of the data points and builds clusters around them~\cite{Mokammel2018,Raatikainen2022}. 
    These clusters and centroids can serve as input for human assessment on identifying and/or confirming relations between requirements grouped into the same cluster (e.g., suggesting requirement relations between cluster centroids and closest data points to each cluster). Figure \ref{fig:kmeans} illustrates the result of running a k-means clustering algorithm on a sub set of requirements from the PURE data set.
    While ``\textit{similar}" relations can be computationally assessed using vector-based similarity metrics, relations with a different semantic (e.g., ``\textit{contradiction}") are challenging from a cognitive point of view for automated analysis. Therefore, for multiclass relation extraction, clustering approaches are limited to the intervention of human assessment.
\end{itemize}

\begin{tcolorbox}
[width=\linewidth, sharp corners=all, colback=white!95!black, boxrule=0.2pt, boxsep=1pt,left=2pt,right=2pt,top=2pt,bottom=2pt]
\includegraphics[height=10pt]{figures/colab.png}
Read and execute \textit{Section 6.1. Supervised Methods} from the Jupyter Notebook for practical application of an LSA-based clustering method to identify related requirements.
\end{tcolorbox} 

\subsubsection{Semi-supervised Methods.} 
\begin{figure*}[htbp]
\centerline{\includegraphics[width=\textwidth]{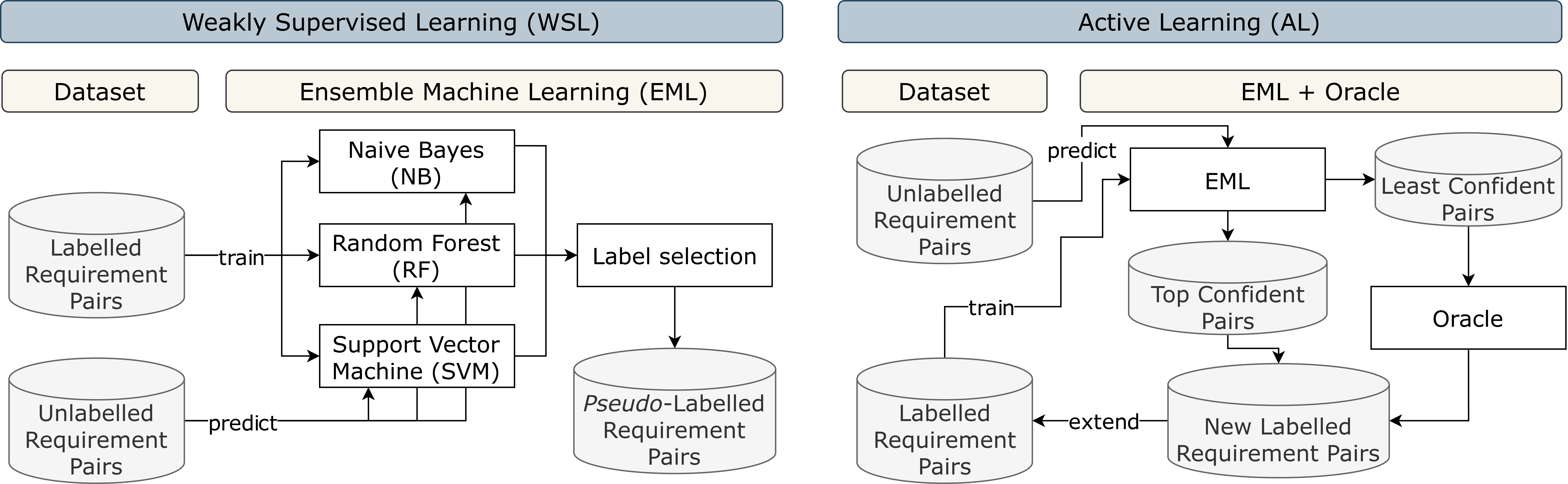}}
\caption{Configuration example of a WSL relation extraction set up using EML (left) and an AL relation extraction set up based on EML (right).}
\label{fig:wsl-al}
\end{figure*}

Combination of labelled and unlabelled data to improve relation extraction. These methods leverage a small amount of labelled relations for training, while using a larger pool of unlabelled data to learn patterns and generalize relationships. Semi-supervised methods reported in Figure \ref{fig:rel-ex} include:

\begin{itemize}

    \item \textbf{Weakly supervised learning (WSL)} Combination of supervised and unsupervised learning to overcome the challenges of limited labelled requirement related pairs. In this context, multiple machine learning models like Naive Bayes (NB), Random Forest (RF) or Support Vector Machines (SVM) can be trained on a small dataset of requirement pairs. These models are then utilized to classify unlabelled data~\cite{Deshpande2019a} and used to identify instances where all classifiers agree on the labelling, considering them as consistent results. This strategy is commonly known as ensemble machine learning (EML). The agreed labels are assigned as pseudo labels to the corresponding data points. By iteratively refining the model using these pseudo labels, the approach enables the extraction of relations from unlabelled data. This WSL-based method showcases how weak supervision can be employed to tackle the challenge of limited labelled data in relation extraction tasks. Figure \ref{fig:wsl-al} illustrates the instantiation of a potential configuration for WSL using EML.
    \item \textbf{Active learning (AL)}. AL approaches design learning algorithms which require from interactive queries to an oracle~\cite{Deshpande2020}, whether it is a human domain expert or a third software component with domain expert knowledge (e.g., an ontology-based dependency prediction tool~\cite{Motger2019}). In this context, an EML learner can be used to classify unlabelled data, but predictions are evaluated based on the prediction confidence level. Uncertain predictions (i.e., below a specific confidence threshold) are sent to an oracle to be labelled accordingly. On the other hand, most confident predictions (i.e., above a specific confidence threshold) are incorporated to the training set. Values for uncertainty thresholds respond to a context-dependent trade-off between accuracy and efficiency of the method. Figure \ref{fig:wsl-al} illustrates the instantiation of a potential configuration for AL using EML.
\end{itemize}

\begin{tcolorbox}
[width=\linewidth, sharp corners=all, colback=white!95!black, boxrule=0.2pt, boxsep=1pt,left=2pt,right=2pt,top=2pt,bottom=2pt]
\includegraphics[height=10pt]{figures/colab.png}
Read and execute \textit{Section 6.2. Semi-supervised Methods} from the Jupyter Notebook for practical application of an ensemble machine learning method using NB, RF and SVM to identify related requirements.
\end{tcolorbox} 

\subsubsection{Supervised Methods.} Based on the evolution and fine-tuning of machine - and deep-learning models using labelled requirement pairs to accurately identify and extract dependencies between requirements. These methods rely on annotated data to learn patterns and linguistic cues associated with different types of requirement relationships, enabling the model to predict and generalize dependencies in new, unseen pairs of requirements. As a generic classification, in Figure \ref{fig:rel-ex} we differentiate between the following:

\begin{itemize}
    \item \textbf{ML classifiers.} As introduced in the description of EML techniques, ML methods employ traditional statistical ML classifiers like Naive Bayes, Support Vector Machine, K-nearest neighbours~\cite{Deshpande2019a,Abeba2022}. Whether in isolation or through their combined use (EML), ML classifiers are used as a binary (type-agnostic) or multi-class (type-dependent) classification algorithm, where models are pre-trained using a data set of labelled requirement pairs $r_i \xrightarrow{d} r_j$ and later used to predict new, unforeseen requirement relations. Input for these classifiers can either be pre-processed natural language representations for each $r_i \in R$~\cite{Deshpande2019}, or vectorization representations using pre-trained models (e.g., Word2Vec, FastText, GloVe)~\cite{Abeba2022}.
    \item \textbf{DL classifiers.} Classification of requirement relations using requirements embeddings as input for deep learning classifiers such as Convolutional Neural Networks (CNN), Long Short-Term Memory networks (LSTM), Bidirectional LSTM (BiLSTM) and Recurrent Neural Networks (RNNs) for relation classification~\cite{Abeba2022,Firmawan2022,Fischbach2021a}, including the use of encoder-based transformer models like BERT~\cite{Fischbach2021,Jadallah2021} for both the generation of requirements embeddings and for fine-tuning LLMs for classification tasks. 
\end{itemize}

\begin{tcolorbox}
[width=\linewidth, sharp corners=all, colback=white!95!black, boxrule=0.2pt, boxsep=1pt,left=2pt,right=2pt,top=2pt,bottom=2pt]
\includegraphics[height=10pt]{figures/colab.png}
Read and execute \textit{Section 6.3. Supervised Methods} from the Jupyter Notebook for practical application of a fine-tuning process of an encoder-only model like BERT with a top layer for document classification to identify related requirements.
\end{tcolorbox}

\section{Comparative analysis}

\subsection{Empirical analysis}
\label{sec:evaluation}

To illustrate the main advantages and drawbacks of retrieval-based and machine learning-based methods respectively, we briefly report on the design principles, effectiveness and performance efficiency of two representative methods. For retrieval-based, we use an ontology-based method using an RDF ontology in OWL format. For machine learning-based, we use a supervised method using a deep learning, Transformer-based classifier.

\subsubsection{Ontology-based: OpenReq ERTMS/ETCS ontology.} We use the OpenReq-DD tool~\cite{Motger2019} as an ontology-based representative, which includes the distribution and evaluation of an ontology in RDF/OWL format within the ERTMS/ETCS domain~\cite{Deshpande2020}. A snapshot of this ontology is presented in Figure~\ref{fig:ontology} (full version is referenced on the Notebook). The complete ontology contains 28 ontology classes, including 25 entity nodes, 3 types of dependencies and 25 dependency instances (7 \textit{requires}, 17 \textit{refines}, 1 \textit{conflicts}). 

\subsubsection{Machine learning-based: fine-tuning BERT with PURE annotated dataset.}

We use a version of BERT (BERT-base-uncased) and we extend it with a fine-tuning process for the sequence classification tasks. In this context, a pair of requirements is used as the full sequence. The fine-tuning process involves the entire PURE annotated dataset, with a 10-fold cross-validation approach (90\% for training, 10\% for testing) to aggregate results and predicted requirement relations.

\subsubsection{Empirical analysis.} Table~\ref{tab:comparative-analysis} reports a summarized overview of the empirical analysis we conducted to compare both strategies, including design specifications, technologies used, retrieved dependencies and execution time. We do not focus on qualitative measures for functional appropriateness (i.e., accuracy, precision, recall, F-measure), as baseline artefacts for ground-truth analysis (i.e., ETC/ERTMS ontology, PURE annotated dataset) are limited and restricted to external constraints (e.g., the ontology does not model \textit{similar} dependencies, while there are no \textit{refines} annotated instances). Furthermore, the purpose of this analysis is to provide a comprehensive overview of the strengths and limitations of each approach, rather than comparing their reliability in a purely academic, illustrative context.

\begin{table}
\caption{Comparative analysis summary of an ontology-based approach vs. a machine learning-based approach. Evaluation is conducted using the PURE annotated dataset for multiclas classification, consisting on 190 unique requirements (see Section~\ref{sec:pure})}
\label{tab:comparative-analysis}
\renewcommand{\arraystretch}{1.2} 
\begin{tabularx}{\textwidth}{p{0.20\textwidth}@{\hspace{8pt}}p{0.37\textwidth}@{\hspace{15pt}}p{0.35\textwidth}}
\toprule
\textbf{Type} & \textbf{Ontology-based} & \textbf{Machine learning-based} \\ \midrule
\textbf{Method} & OpenReq-DD with domain-specific RDF/OWL ontology & Fine-tuned BERT instance with annotated PURE dataset \\
\textbf{Design} & Ontology generation (5 hours) & -- \\
\textbf{Technologies} & OWL/RDF (ontology design) WordNet (similarity analysis) & BERT (seq. classification) \\
\textbf{NLP Prep.} & \multicolumn{2}{m{0.76\textwidth}}{Transformer-based tokenizer, Attribute ruler, Lemmatizer, Part-of-Speech tagger, N-gram generation} \\
\hline
\textbf{\#non-relevant} & 16,969 pairs & 16,558 pairs \\
\textbf{\#refines} & 307 pairs& -- \\
\textbf{\#requires} & 679 pairs& 721 pairs \\
\textbf{\#similar} & -- & 676 pairs \\
\hline
\textbf{Ex. Time} & 27 seconds & 9,224 seconds (model training) 381 seconds (inference)\\
\hline
\end{tabularx}
\end{table}

In terms of NLP-based knowledge representation, both approaches utilize the same subset of syntactic techniques (tokenization, lemmatization, PoS tagging, n-gram generation). Notably, the ontology-based method requires ontology generation, which in this context took approximately 5 hours~\cite{Deshpande2020}, highlighting a dependence on expert knowledge and conceptual modeling, a characteristic not shared by the machine learning-based method. In terms of requirement dependencies, the ontology-based approach identifies 307 \textit{refines} and 679 \textit{requires} instances, while the machine learning-based approach detects 721 \textit{requires} pairs and 676 \textit{similar} instances. This discrepancy underscores the nuanced nature of relations captured by each method, limited by either the ontology knowledge or the annotated dataset. Regarding execution time, the ontology-based approach exhibits efficiency, completing the extraction process in 27 seconds, while the machine learning-based method involves a training duration of 9,224 seconds (\~2.5 hours) and an inference time of 381 seconds (\~6 minutes) for all potentially dependent requirement pairs. 

\subsection{Discussion}
\label{sec:discussion}

The selection of an appropriate relation extraction method is influenced by multiple contextual factors (domain, data, expert knowledge, computational resources, skills...). To guide this selection process, we use the empirical analysis in Section~\ref{sec:evaluation} as a trigger for the elicitation of the strengths and limitations of retrieval-based and machine learning-based strategies.

Retrieval-based relation extraction methods present the following strengths with respect to machine learning methods:

\begin{itemize}
    \item \textbf{Interpretability.} Retrieval-based methods are relatively simple to implement and understand, with respect to the theoretical foundations of syntactic and semantic relations between requirements. On the other hand, interpretability of deep learning models can be limited, making it difficult to understand how specific relations are extracted.
    \item \textbf{Representativeness.} Retrieval-based systems can be used to extract a wide, customized range of relations, while machine learning approaches may suffer from limited representativeness if the data is biased or incomplete.
    \item \textbf{Data availability.} They can be used without annotated requirement relation instances, whereas machine learning methods require substantial labelling efforts to create training datasets.
    \item \textbf{Computational resources.} Retrieval-based approaches are typically low consuming in terms of computational resources, while some deep learning models may be computationally expensive and require powerful hardware for training and inference.
\end{itemize}

On the other hand, machine learning methods present the following strengths:

\begin{itemize}
    \item \textbf{Expressive power.} Machine learning methods can handle complex patterns and relationships in requirements text, whereas retrieval-based methods can be brittle, meaning that they may not be able to handle unexpected or unusual text, including format and content.
    \item \textbf{Adaptability.} Machine learning methods can automatically learn from data and adapt to new patterns without the need for explicit human-defined rules, making them more flexible. On the other hand, retrieval-based approaches can be difficult to maintain, as they require the rules to be updated as the requirements and relations change.
    \item \textbf{Scalability.} Machine learning methods can typically handle a large amount of data efficiently, enabling scalability to big datasets, whereas retrieval-based approaches might not always adapt to large-scale contexts.
    \item \textbf{Required knowledge.} Deep learning models used in machine learning approaches can capture intricate semantic relationships, whereas retrieval-based methods typically require human expert knowledge for conceptual modelling of relations (e.g., ontologies, patterns, rules, dictionaries).
\end{itemize}
\section{Challenges and Future Directions}
\label{sec:challenges}

Among the most significant limitations of NLP-based relation extraction methods, and triggered by the surveyed literature in this chapter, we highlight the following:

\begin{itemize}
    \item \textbf{Lack of annotated data.} One of the major challenges in applying NLP techniques for relation extraction in RE is the scarcity of labelled data. Supervised machine learning methods heavily rely on annotated requirement pairs for training, and obtaining a sufficient amount of accurately labelled data is often a time-consuming and costly process.
    \item \textbf{Not uniform categorization of relations.} Defining a standardized and uniform categorization of requirement relations remains a significant challenge. In real-world scenarios, different projects and domains might require distinct relation types, making it difficult to generalize extraction methods across diverse RE contexts.
    \item \textbf{Modelling human knowledge (semantics of relation types).} Capturing and incorporating domain-specific human knowledge and semantics related to dependency types into automated extraction techniques is complex. Retrieval-based and machine learning-based methods might struggle to fully comprehend and accurately model the intricacies of such knowledge.
\end{itemize}

To address the limitations and challenges mentioned above, state-of-the-art solutions in the NLP4RE field focus on the following directions:

\begin{itemize}
    \item \textbf{Extend contributions using encoder-based LLMs.} Recent advancements in encoder-based large language models, such as BERT, have shown promising results in various NLP4RE tasks, including requirements extraction~\cite{DeAraujo20211321}, classification~\cite{Hey2020169} and disambiguation~\cite{Ezzini2022187}, among others. Extending the application of these models to relation extraction in RE can leverage their contextual understanding and semantic representation capabilities to improve accuracy and adaptability, even in data-scarce scenarios.
    \item \textbf{Explore generative LLMs for zero-shot or few-shot relation extraction.} Generative language models such as GPT-4~\cite{openai2023gpt4}, LLaMA~\cite{touvron2023llama} and PaLM~\cite{chowdhery2022palm} have demonstrated their ability to produce coherent and contextually relevant text. In the field of NLP4RE, these models are being used for requirements generation and augmentation~\cite{Grasler2022} and user analysis~\cite{Clements2023701}, among others. 
    Investigating their potential for generating relation instances between requirements can assist in building larger labelled datasets and address the issue of data scarcity in supervised learning.
\end{itemize}
\section{Summary and Conclusion}
\label{sec:conclusions}

The fusion of NLP and RE has unveiled innovative pathways for the automated extraction of requirement relations. From purely retrieval-based to machine learning-based approaches, each with distinct advantages, these methods lay the groundwork to support the extraction and categorization of dependencies in specialized, domain-specific domains. Retrieval-based methods offer simplicity through linguistic patterns and predefined ontologies, while machine learning-based approaches harness language models for increased adaptability and efficiency in large scale domains.

However, several challenges persist, such as the scarcity of annotated data and the absence of standardized relation categorization. Encouragingly, encoder-based large language models like BERT hold promise for enhancing extraction accuracy, even with limited data (i.e., few-shot learning). Additionally, generative models offer great potential in generating relation instances to enrich datasets or even assisting in the relation extraction process itself.

These methodologies not only deepen our understanding of the landscape of NLP-based relation extraction, but also inspire future innovations that bridge human expertise with automated analysis. The collaboration between linguistic insights and computational advancements paves the way for improved decision-making, efficient requirement management, and enhanced system design. All in all, this chapter aims at encouraging researchers and practitioners to further contribute to the progression of automated relation extraction in the domain of NLP4RE.
\section*{Acknowledgements}
\label{sec:acks}

With the support from the Secretariat for Universities and Research of the Ministry of Business and Knowledge of the Government of Catalonia and the European Social Fund.

\bibliographystyle{splncs04}
\bibliography{bib}
\end{document}